# Comparison of Light-Time Formulations in the post-Newtonian framework for the BepiColombo MORE experiment

P. Cappuccio[1], I. di Stefano[1], G. Cascioli[1] and L. Iess[1]

**Abstract**

The ESA/JAXA BepiColombo mission, launched on 20 October 2018, is currently in cruise towards Mercury. The Mercury Orbiter Radio-science Experiment (MORE), one of the 16 experiments of the mission, will exploit range and range-rate measurements collected during superior solar conjunctions to better constrain the post-Newtonian parameter $\gamma$. The MORE radio tracking system is capable of establishing a 5-leg link in X- and Ka-band to obtain 2-way range-rate measurements with an accuracy of 0.01 mm/s @ 60 s sampling time and 2-way range measurements at centimeter level after a few seconds of integration time, at almost all solar elongation angles. In this paper, we investigate if the light-time formulation derived by T. Moyer, implemented in JPL's orbit determination code MONTE, is still a valid approximation, in light of the recent advancements in radiometric measurement performance. Several formulations of the gravitational time delay, expressed as an expansion in powers of $GM/c^2r$, are considered in this work. We quantified the contribution of each term of the light-time expansion for the first superior solar conjunction experiment of BepiColombo. The maximum 2-way error caused by Moyer approximation with respect to a complete second order expansion amounts to 17 mm. This is at the level of accuracy of the novel pseudo-noise (PN) ranging system at 24 Mcps used by MORE. A complete second order expansion is then recommended for present and future superior solar conjunction experiments. The perturbation caused by the planets in the solar system is considered as well, resulting in significant effects due to the Jupiter, the Earth and the Saturn systems. For these bodies the classical Shapiro time delay is sufficient. The corrections due to the Sun oblateness and angular momentum are negligible. The aforementioned considerations are valid for all superior conjunction experiments involving state-of-the-art radio-tracking measurements.

---

[1] Department of Mechanical and Space Engineering, Sapienza University of Rome, Via Eudossiana 18, Rome, Italy.

1. Introduction

The parameterized post-Newtonian (PPN) framework was introduced by Eddington back in 1922 to identify possible deviations from general relativity in a quasi-stationary weak field. In his textbook *The Mathematical Theory of General Relativity,* Eddington introduced 3 parameters, which become 10 in the modern version of the PPN framework by Nordtvedt and Will [1]–[3]. These 10 parameters are related with different physical phenomena, including the amount of space curvature produced by a rest mass, the non-linearity in the superposition of gravity, the conservation of total momentum, preferred-location and preferred-frame effects [4]. In this work, we are interested only in the parameter $\gamma$, also called Eddington parameter, that controls the deflection of light, the delay and the Doppler shift of photons propagating near a massive body.

The first approximation of the general relativistic time delay perturbation was formulated by Shapiro in 1964 [5]. He considered first-order expansion in powers of $GM_\odot/c^2 r$, which was sufficiently accurate for the tests of relativistic time delay performed at that time [6]. Developments of measurement systems required more accurate formulations with an expansion at least to the second order in $GM_\odot/c^2 r$. Hitherto, the most accurate estimation of the Eddington parameter was provided by the Cassini superior solar conjunction experiment in 2002 [7]. The geometry of a superior solar conjunction is reported in Fig. 1. The Cassini experiment provided an estimation of $\gamma = 1 + (2.1 \pm 2.3) \times 10^{-5}$ using a Doppler dataset obtained through a 5-leg multifrequency link. The link was composed of a X-band uplink coherent with two X- and Ka-band downlinks, and a Ka-band uplink coherent with a Ka-band downlink. This setup allowed to obtain plasma-free 2-way range-rate with a 1-way accuracy of $2.2 \cdot 10^{-6} \ m/s$ at 300 s of integration time, when the impact parameter (the distance between the signal-path and the center of mass of the Sun) was above 7 solar radii.

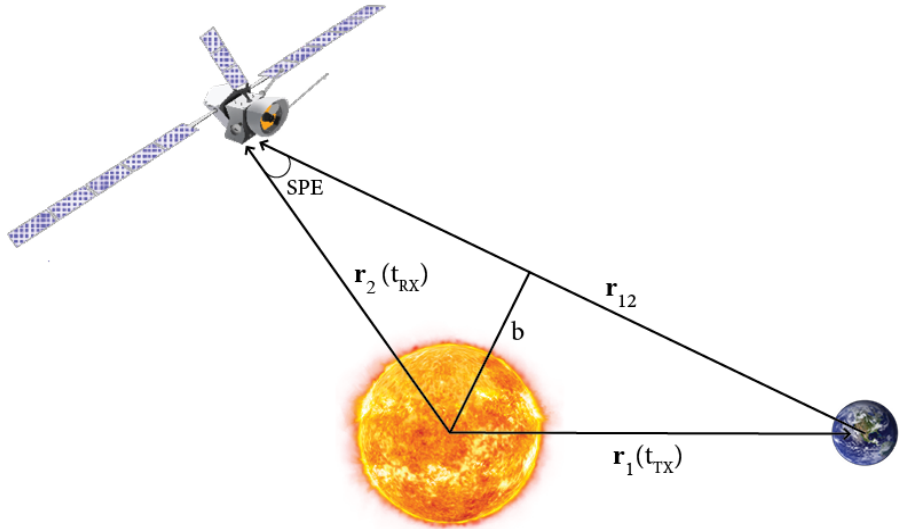

Figure 1. Geometry of a superior conjunction experiment. The impact parameter, b, is the distance between the light-path and the center of mass of the perturbing mass (i.e., the Sun). $r_1$ is the distance of the transmitter from the Sun center of mass at the transmit time, $r_2$ is the distance of the receiver from the Sun center of mass at the receiving time and $r_{12}$ is the distance between the receiver at the receiving time and the transmitter at the transmitting time.

The next superior solar conjunction experiment intended to improve our knowledge of the Eddington parameter is performed in the context of the Mercury Orbiter Radio-science Experiment (MORE) [8] of the BepiColombo mission. The MORE experiment relies on a radio-tracking system similar to the one of Cassini (5-leg radio link in X- and Ka-band) [9]. In addition, MORE uses a state-of-the-art Pseudo-Noise (PN) ranging system. In particular, the MORE Ka-band transponder (KaT) provides 24 Mcps PN range measurements with cm-level accuracy [10] and range-rate with an accuracy comparable to the one of Cassini. Numerical simulations of superior solar conjunction experiments during the cruise phase of BepiColombo show that MORE will be able to estimate $\gamma$ at about $6 \cdot 10^{-6}$ ([11], [12]). Thanks to the additional SSCs occurring during the Hermean phase, MORE is expected to estimate $\gamma$ at the level of $2 \cdot 10^{-6}$, together with several PPN parameters [13], [14]. In addition to fundamental physics test, MORE will carry out a thorough investigation of Mercury geodesy and geophysics by accurately estimating the planet gravitational field and rotational state [15]–[17].

In [7], Cassini data were analyzed by using the JPL's Orbit Determination Program (ODP) [18], a spacecraft navigation code that implemented an approximation of the light-time formulation of order $(GM_\odot/c^2r)^2$. The first BepiColombo superior solar conjunction, which took place from 10 to 24 March 2021, is used as a benchmark to evaluate if the terms neglected in the ODP are relevant for the MORE experiment. To analyze MORE data, we will use the JPL's Mission Analysis, Operations, and Navigation Toolkit Environment (MONTE) [19], a modern version of the ODP. While it has been proven to be accurate enough for the Cassini experiment ([20], [21]), this may not be the case for BepiColombo. In this paper, we investigate whether new advancements in radio-tracking systems accuracies require an improvement of the modelling of the light-time in the orbit determination software.

The paper is organized as follows: Sec. 2 introduces the different light-time formulations; Sec. 3 presents the method used to carry out the analysis; Sec. 4, shows results of the analysis and finally, Sec. 5 provides the conclusions of this work.

## 2. Light-Time Formulations

In the case of a static and spherically symmetric body, neglecting terms of order higher than $(GM_\odot/c^2r)^2$, the metric tensor can be written as [22]:

$$ds^2 = \left(1 - \frac{2GM}{c^2r} + 2\beta\frac{G^2M^2}{c^4r^2}\right)dt^2 - \left(1 + \gamma\frac{2GM}{c^2r} + \frac{3}{2}\epsilon\frac{G^2M^2}{c^4r^2}\right)dl^2 \qquad (1)$$

where $c$ is the speed of light, $G$ is the gravitational constant, $M$ is the mass of the perturbing body, $r$ is the distance of the test mass from the perturbing body, $dl^2$ is the Euclidean line element and $t$ is the Killing time. $\gamma$ and $\beta$ are the post-Newtonian parameters relevant in this metric, which in general relativity are equal to 1. $\epsilon$ is a post-post-Newtonian term [23], not present in the original PPN formalism, again unity in general relativity. In generic scalar-tensor theories the parameter $\epsilon$ depends linearly on $\gamma$ and $\beta$, therefore becoming redundant, but in scalar-tensor-vector theories of gravity this is not true ([24], [25]). Note that Teyssandier and le Poncin-Lafitte [26], in eq. 57, use an equivalent form of the metric, where the post-post-Newtonian parameter $\epsilon$ is referred to as $\delta$.

From the above-mentioned metric, it is possible to compute the time, $\Delta t$, that a photon takes to travel from the transmitter, $r_1$, to the receiver, $r_2$. Several equivalent derivations of the light-time are present in literature ([22], [26]–[28]). With respect to the above-mentioned formulations, Kopeikin [28] also includes a Lorentz transformation to convert the light-time from the heliocentric reference system to the to the solar system barycenter reference system. Moyer [18] approximates the exact formulation of the light-time up to order $(GM_\odot/c^2)^2$, which he simplified to be accurate enough in all practical cases of space missions, equipped with the most accurate tracking system at that time. All the relevant formulations are listed below, using a common formalism. The classical Shapiro time delay is indicated as $\Delta t_{1PN}$. We refer to the eq. 8.54 of Moyer [18] with $\Delta t_{ODP}$, and to eq. 94 of [22], or equivalently to eq. 67 of [26], with $\Delta t_{2PN}$. We refer to eq. 71 of [28] as $\Delta t_{SSB}$ (with SSB standing for solar system barycenter).

$$\Delta t_{1PN} = \frac{|r_{12}^{SSB}|}{c} + \frac{(1+\gamma)GM_\odot}{c^3} \ln\left(\frac{|r_2|+|r_1|+|r_{12}|}{|r_2|+|r_1|-|r_{12}|}\right) \tag{2}$$

$$\Delta t_{ODP} = \frac{|r_{12}^{SSB}|}{c} + \frac{(1+\gamma)GM_\odot}{c^3} \ln\left(\frac{|r_2|+|r_1|+|r_{12}|+\frac{(1+\gamma)GM_\odot}{c^2}}{|r_2|+|r_1|-|r_{12}|+\frac{(1+\gamma)GM_\odot}{c^2}}\right) \tag{3}$$

$$\Delta t_{2PN} = \frac{|r_{12}^{SSB}|}{c} + \frac{(1+\gamma)GM_\odot}{c^3} \ln\left(\frac{|r_2|+|r_1|+|r_{12}|}{|r_2|+|r_1|-|r_{12}|}\right) +$$
$$+ \frac{G^2 M_\odot^2}{c^5} \frac{|r_{12}|}{|r_2||r_1|} \left[\left(\frac{8(1+\gamma)-4\beta+3\epsilon}{4}\right) \frac{\arccos(\hat{r}_1 \cdot \hat{r}_2)}{|\hat{r}_1 \times \hat{r}_2|} - \frac{(\gamma+1)^2}{1+\hat{r}_1 \cdot \hat{r}_2}\right] \tag{4}$$

$$\Delta t_{SSB} = \frac{|\boldsymbol{r}_{12}^{SSB}|}{c} + \left[(1+\gamma)\left(1 - \hat{\boldsymbol{r}}_{12} \cdot \frac{\boldsymbol{v}_\odot}{c}\right) + \frac{1}{2}\left(\frac{v_\odot}{c}\right)^2\right]\frac{GM_\odot}{c^3}\ln\left(\frac{|\boldsymbol{r}_2| + |\boldsymbol{r}_1| + |\boldsymbol{r}_{12}|}{|\boldsymbol{r}_2| + |\boldsymbol{r}_1| - |\boldsymbol{r}_{12}|}\right) +$$
$$+ \frac{G^2 M_\odot^2}{c^5} \frac{|\boldsymbol{r}_{12}|}{|\boldsymbol{r}_2||\boldsymbol{r}_1|}\left[\left(\frac{8(1+\gamma) - 4\beta + 3\epsilon}{4}\right)\frac{\arccos(\hat{\boldsymbol{r}}_1 \cdot \hat{\boldsymbol{r}}_2)}{|\hat{\boldsymbol{r}}_1 \times \hat{\boldsymbol{r}}_2|} - \frac{(\gamma+1)^2}{1 + \hat{\boldsymbol{r}}_1 \cdot \hat{\boldsymbol{r}}_2}\right] +$$
$$+ \frac{\gamma+1}{2}\frac{GM_\odot}{c^3}\frac{|\boldsymbol{r}_{12}|}{|\boldsymbol{r}_2||\boldsymbol{r}_1|}\frac{\left(\hat{\boldsymbol{r}}_1 \times \frac{\boldsymbol{v}_\odot}{c}\right)^2 |\boldsymbol{r}_1| + \left(\hat{\boldsymbol{r}}_2 \times \frac{\boldsymbol{v}_\odot}{c}\right)^2 |\boldsymbol{r}_2| - \left(\hat{\boldsymbol{r}}_{12} \times \frac{\boldsymbol{v}_\odot}{c}\right)^2 (|\boldsymbol{r}_2| + |\boldsymbol{r}_1|)}{1 + \hat{\boldsymbol{r}}_1 \cdot \hat{\boldsymbol{r}}_2} \quad (5)$$

$$\boldsymbol{r}_{12} = \boldsymbol{r}_2(t_{RX}) - \boldsymbol{r}_1(t_{TX}) \quad (6)$$

$\boldsymbol{r}_1$ is the transmitter vector with respect to the Sun center of mass at the transmitting time, $t_{TX}$; $\boldsymbol{r}_2$ is the receiver vector with respect to the Sun center of mass at the receiving time, $t_{RX}$, and $\boldsymbol{v}_\odot$ is velocity of the Sun with respect to the solar system barycenter. $\boldsymbol{r}_{12}^{SSB}$ is the vector from the transmitter at $t_{TX}$ to the receiver at $t_{RX}$ computed with respect to the solar system barycenter, which represents the geometrical distance between the two points.

Eq. 8 and 9 report the upper limit for the $\delta t_{\Delta pN}$ and the $\delta t_{ppN}$ terms, as reported in eq. 25 and 26 of [29]; eq. 7 and 10 provide an approximation to the upper limit for the $\delta t_{1PN}$ and the $\delta t_{SSB}$ term [4]:

$$|\delta t_{1PN}| = \left|\frac{(1+\gamma)GM_\odot}{c^3}\ln\left(\frac{|\boldsymbol{r}_2| + |\boldsymbol{r}_1| + |\boldsymbol{r}_{12}|}{|\boldsymbol{r}_2| + |\boldsymbol{r}_1| - |\boldsymbol{r}_{12}|}\right)\right| \leq \frac{(1+\gamma)GM_\odot}{c^3}\ln\left(\frac{|\boldsymbol{r}_2||\boldsymbol{r}_1|}{b_{min}^2}\right) \quad (7)$$

$$|\delta t_{\Delta pN}| = \left|\frac{G^2 M_\odot^2}{c^5}\frac{|\boldsymbol{r}_{12}|}{|\boldsymbol{r}_2||\boldsymbol{r}_1|}\frac{(\gamma+1)^2}{1+\hat{\boldsymbol{r}}_1 \cdot \hat{\boldsymbol{r}}_2}\right| \leq \frac{2G^2 M_\odot^2}{c^5 b_{min}^2}|\boldsymbol{r}_{12}| \quad (8)$$

$$|\delta t_{ppN}| = \left|\frac{G^2 M_\odot^2}{c^5}\frac{|\boldsymbol{r}_{12}|}{|\boldsymbol{r}_2||\boldsymbol{r}_1|}\left(\frac{8(1+\gamma) - 4\beta + 3\epsilon}{4}\right)\frac{\arccos(\hat{\boldsymbol{r}}_1 \cdot \hat{\boldsymbol{r}}_2)}{|\hat{\boldsymbol{r}}_1 \times \hat{\boldsymbol{r}}_2|}\right| \leq \frac{15}{4}\pi\frac{G^2 M_\odot^2}{c^5 b_{min}} \quad (9)$$

$$|\delta t_{SSB}| = \left|\left[(1+\gamma)\left(-\hat{\boldsymbol{r}}_{12} \cdot \frac{\boldsymbol{v}_\odot}{c}\right) + \left(\frac{v_\odot}{c}\right)^2\right]\frac{GM_\odot}{c^3}\ln\left(\frac{|\boldsymbol{r}_2| + |\boldsymbol{r}_1| + |\boldsymbol{r}_{12}|}{|\boldsymbol{r}_2| + |\boldsymbol{r}_1| - |\boldsymbol{r}_{12}|}\right)\right| \leq \frac{v_\odot}{c}\delta t_{1PN} \quad (10)$$

where $b_{min}$ is the minimum value of the impact parameter, i.e., the minimum distance between the light-path and the center of mass of the Sun. The above upper limits (eq. 7 to 10) can be used as a first evaluation for the different contributions during a conjunction. $\delta t_{1PN}$ is the first order relativistic correction to the light-time due to the geometrical distance. $\delta t_{\Delta pN}$ is the second order correction that provides the largest contribution during a superior conjunction. Indeed, it is known as the *enhanced*

*term* [22]. The Moyer approximation takes into account only this second order term. This term is accounted for by the term $\frac{(1+\gamma)GM_\odot}{c^2}$ inside the logarithm (see chapter 8 of [18] for its mathematical derivation). $\delta t_{ppN}$ is a post-post-Newtonian second order correction which has a smaller effect during a superior conjunction. It is neglected by Moyer. $\delta t_{SSB}$ corresponds to a Shapiro time delay which accounts also for a Lorentz transformation from the Sun space-time reference frame to the solar system barycenter reference frame. It takes explicitly into account the fact that the Sun moves with a non-null velocity with respect to the solar system barycenter. The Moyer implementation of $\Delta t_{ODP}$ implicitly considers the motion of the Sun by computing the instantaneous value of $r_1$ and $r_2$ using Sun ephemerides (i.e., in the computation of $r_1$ and $r_2$, the position of the Sun with respect to the solar system barycenter at $t_{TX}$ is not equal to the position at $t_{RX}$).

In addition to PPN corrections of the light-time induced by a static and spherically symmetric body, we should also consider the delay due to the oblateness and rotation of the massive body.

The effect of the oblateness of an axisymmetric body can be written as eq. 11 [23], [30]–[32]. The gravitomagnetic correction due to the Sun angular momentum can be expressed as eq. 12 (from eq. 16 of [33]).

$$|\delta t_{J_2}| = \left| \frac{(1+\gamma)GM_\odot}{2c^3} \frac{J_{2\odot} R_\odot^2}{|r_2||r_1|} \frac{|r_{12}|}{1+\hat{r}_1 \cdot \hat{r}_2} \times \left[ \frac{1-(\hat{k}_\odot \cdot \hat{r}_1)^2}{|r_1|} + \frac{1-(\hat{k}_\odot \cdot \hat{r}_2)^2}{|r_2|} - \left(\frac{1}{|r_1|} + \frac{1}{|r_2|}\right) \frac{[\hat{k}_\odot \cdot (\hat{r}_1 + \hat{r}_2)]^2}{1+\hat{r}_1 \cdot \hat{r}_2} \right] \right|$$
$$\leq \frac{(1+\gamma)GM_\odot}{c^3} \left(\frac{R_\odot}{b}\right)^2 J_{2\odot} \qquad (11)$$

$$|\delta t_{AM}| = \left| -\frac{(1+\gamma)GS_\odot}{c^4} \left(\frac{1}{|r_1|} + \frac{1}{|r_2|}\right) \frac{\hat{k}_\odot \cdot (\hat{r}_1 \times \hat{r}_2)}{1+\hat{r}_1 \cdot \hat{r}_2} \right| \leq \left| -\frac{4GS_\odot}{c^4 b} \right| \qquad (12)$$

Where $J_{2\odot}$ is the quadrupole term of the spherical harmonic expansion of the Sun gravity field, $R_\odot$ is its equatorial radius, $\hat{k}_\odot$ is the unit vector of pole direction in the inertial reference frame, and $S_\odot$ is the angular momentum of the Sun. The effect of $J_2$, eq .11) is already implemented in MONTE.

3. **Method**

The synthetic dataset of the first superior solar conjunction (10-24 March 2021) of BepiColombo is used as a benchmark to identify the differences between the four formulations (eq. 2 to 5). However, our results are applicable not only for the BepiColombo mission, but, more broadly to all superior conjunction experiments in the solar system relying on current state-of-the-art microwave tracking systems. To test the different $\Delta t$ we used the latest release of the official BepiColombo kernel [34] and the JPL DE432 planetary ephemerides, which are based on the DE430 [35]. From this dataset, we extracted all the necessary quantities: $\boldsymbol{r}_1, \boldsymbol{r}_2, GM_\odot$, and $\boldsymbol{v}_\odot$.

The value of the Sun gravitational constant, $GM_\odot$, is equal to $1.327124400419394 \cdot 10^{11} \, km^3/s^2$ [35]. The most recent release of the JPL ephemerides [36] reports a slightly different value for the gravitational parameter of the Sun, but this difference is not relevant to our analysis. The average velocity of the Sun with respect to the solar system barycenter is ~ 15.6 m/s. The minimum value assumed by the impact parameter during the first superior solar conjunction of BepiColombo is equal to 4.3 $R_\odot$ (~ $3 \cdot 10^6 \, km$). Here, we consider the data simulated at all values of the impact parameter, even if Cassini data analysis has shown that measurements collected when $b$ is below 7-8 solar radii must be discarded because the X-band link enters in strong scintillation regime, hindering the application of the dispersive noise cancellation scheme [37].

We verified through numerical analysis that $\Delta t_{ODP}$ (eq. 2) is the formulation used by JPL's MONTE navigation toolkit [19]. This is the same formulation used for the Cassini solar conjunction experiment [7]. In what follows, referring to Moyer will be the same as referring to the ODP and MONTE formulation.

MONTE offers the possibility to compute the time delay due to multiple bodies at the same time. For bodies other than the Sun, Moyer computes only the $\delta t_{1PN}$ and sums linearly the contribution for each body (formula 8.55 of [18]). The relativistic light-time due to each body is calculated in the space-time reference frame of that body. Eq. 10 provides the error related to the absence of the Lorentz transformation to refer the light-time from the perturbing body space-time frame to the solar system barycenter.

We used the Sun gravitational moment $J_{2\odot} = 2.246 \cdot 10^{-7}$ from the latest estimate of the MESSENGER mission data [38], which is accordance with helioseismology models [39], [40] and ephemerides estimation [41]. We used the value of the Sun angular momentum, $S_\odot = 1.92 \cdot 10^{41}\ kg\ m^2\ s^{-1}$, obtained by helioseismology [42].

## 4. Results

We express the results in units of spatial length to provide an easier comparison with the state-of-the-art performance of radio-tracking systems (~1 cm) and a more tangible interpretation. We performed a series of comparisons to characterize the effect of each term in the relativistic formulation.

Fig. 2 shows the Shapiro time delay on the BepiColombo-Earth 2-way link during the first superior solar conjunction once the geometrical distance, $|\mathbf{r}_{12}^{SSB}|/c$, is removed (i.e. $\delta t_{1PN}$). The signal reaches a peak value of 50.6 km when $b$ is at its minimum.

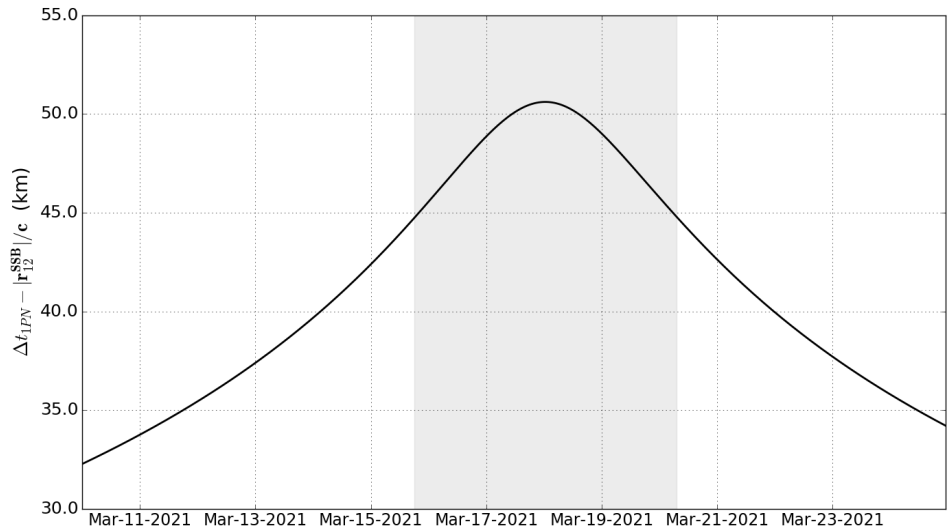

Figure 2. First-order 2-way relativistic time delay correction, $\delta t_{1PN}$, on the first BepiColombo superior solar conjunction. The peak value, occurring when the impact parameter is at its minimum, is 50.6 km. The gray shaded area indicates that $b$ is below 7 solar radii.

The top panel of Fig 3. shows the effect of the second order term enhanced during a superior solar conjunction, which is the approximation of $\delta t_{\Delta pN}$ used by Moyer. This correction reaches a maximum value of about -202.5 mm at its peak and it is $4 \times 10^{-6}\ \delta t_{1PN}$. The middle panel of Fig.3 reports the comparison between the full second order relativistic correction, eq. 4, and the MONTE implementation. The maximum difference amounts to 17 mm. This is due to the post-post-Newtonian term, $\delta t_{ppN}$, not considered in the Moyer approximation. Comparing the top and middle panels, it is clear that $\delta t_{\Delta pN}$ reduces the value of the total light-time, which is increased by $\delta t_{ppN}$. These effects can be summed up linearly, thus providing a total value at peak of about -185.5 mm ($\Delta t_{2PN} - \Delta t_{1PN}$), instead of -202.5 mm.

The bottom panel of Fig.3 reports the relativistic light-time perturbation due to the explicit correction for the motion of the Sun with respect to the solar system barycenter, $\delta t_{SSB}$. It consists of a Lorentz transformation from the Sun to the solar system barycenter. The variation on the downlink leg cancels out with the variation on the uplink leg because $v_\odot$ (almost constant) is projected on the unit vectors $\hat{r}_{12}^{uplink}$ and $\hat{r}_{12}^{downlink}$, which are almost parallel but have opposite orientation (i.e., $\hat{r}_{12}^{uplink} \approx -\hat{r}_{12}^{downlink}$). The perturbation on the single uplink or downlink leg is marginal (<0.9 mm), while the overall effect on the 2-way link is negligible (<0.001 mm).

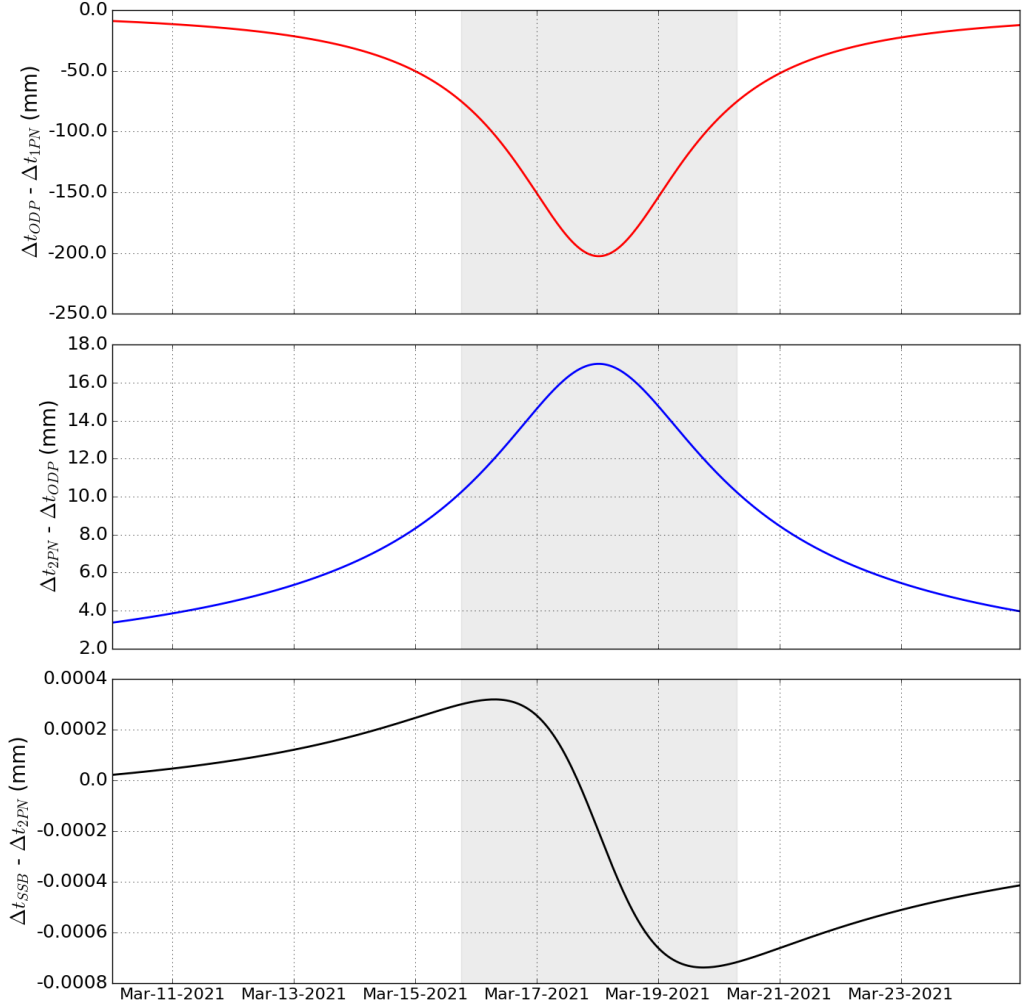

Figure 3. Second order 2-way relativistic light-time correction for the first BepiColombo superior solar conjunction. The red line is obtained by subtracting $\Delta t_{1PN}$ from $\Delta t_{ODP}$, the blue line represents the difference between $\Delta t_{2PN}$ and $\Delta t_{ODP}$ and the black line is the 2-way correction due to the motion of the Sun with respect to the solar system barycenter, $\delta t_{SSB}$. The maximum value of $\delta t_{SSB}$ on the uplink leg is below 0.9 mm and it is compensated by the downlink leg correction (see text). The gray shaded area shows when $b < 7$ solar radii.

Eq. 3, adopted in MONTE, embodies only an approximation of the second-order term which is enhanced during a superior conjunction. To evaluate at which level $\delta t_{\Delta pN}$ is well approximated by the Moyer implementation, $\Delta t_{ODP}$ (eq. 3) is subtracted from $\Delta t_{2PN\_mod}$ (eq. 13). The latter is obtained from eq. 4 by removing $\delta t_{ppN}$, i.e. the term in *arccos*:

$$\Delta t_{2PN\_mod} = \frac{|\mathbf{r}_{12}^{SSB}|}{c} + \frac{(1+\gamma)GM_\odot}{c^3}\ln\left(\frac{|\mathbf{r}_2|+|\mathbf{r}_1|+|\mathbf{r}_{12}|}{|\mathbf{r}_2|+|\mathbf{r}_1|-|\mathbf{r}_{12}|}\right) + $$
$$+ \frac{G^2 M_\odot^2}{c^5}\frac{|\mathbf{r}_{12}|}{|\mathbf{r}_2||\mathbf{r}_1|}\left[-\frac{(\gamma+1)^2}{1+\hat{\mathbf{r}}_1\cdot\hat{\mathbf{r}}_2}\right] \tag{13}$$

A numerical test shows that mismodelling associated to this approximation is negligible (peak error < 0.0035 mm), so that the Moyer approximation of the enhanced term can be used throughout a solar conjunction for missions equipped with standard or Cassini-like radio systems. Eq. 3 still misses the term $\delta t_{ppN}$, which amounts to a maximum of about 17 mm (blue line in fig. 3).

The $\delta t_{J_2}$ on the 2-way measurement has a maximum value of ~0.07 mm. The 1-way gravitomagnetic correction due to the Sun angular momentum, $\delta t_{AM}$, is about 0.32 mm at peak. Due to the chirality of this phenomenon, the 2-way effect is about 0.0033 mm since the uplink and downlink contribution almost cancel out each other. These terms are negligible for our purposes.

Table. 1 reports a comparison of the significant contributions to the relativistic light-time together with their peak value during the first superior solar conjunction of BepiColombo.

Table 1. Effect of 2-way relativistic correction of the light-time applied to the BepiColombo first superior solar conjunction.

| Effect | Formula | Maximum value of relativistic correction |
|---|---|---|
| **Shapiro delay** | $\Delta t_{1PN} - \dfrac{|r_{12}^{SSB}|}{c}$ | 50.6 km |
| **Approximated second order correction** | $\Delta t_{ODP} - \Delta t_{1PN}$ | -202.5 mm |
| **Second order term neglected by Moyer** | $\Delta t_{2PN} - \Delta t_{ODP}$ | 17 mm |
| **Sun oblateness** | $\delta t_{J_2}$ | 0.07 mm |
| **Sun angular momentum** | $\delta t_{AM}$ | 0.003 mm |

Furthermore, we evaluated whether the differences between the formulations vary by perturbing the value of $\gamma$. We found that the deviations are well below relativity experiment sensitivity. As an example, the maximum value of the difference between $\Delta t_{2PN} - \Delta t_{ODP}$ with $\gamma = 1 - 10^{-5}$ and

$\Delta t_{2PN} - \Delta t_{ODP}$ with $\gamma = 1$ amounts to ~90 nm. For our purposes, it means that the effect of the mismodelling of the light-time is independent from the value of $\gamma$.

The Cassini experiment obtained a $\sigma_\gamma = 2.3 \cdot 10^{-5}$, which can be converted to an accuracy on the 1-way light-time of ~ 30 cm [20]. As confirmed by previous works (see Sec. 1), we conclude that the Moyer implementation was accurate enough for the Cassini experiment. During the cruise phase, MORE is expected to be able to estimate $\gamma$ at about $6 \cdot 10^{-6}$, which corresponds to ~ 15 cm, 2-way. Even if this value is about one order of magnitude larger than the unmodelled terms in the MONTE formulation, the fit of the accurate radiometric data acquired near the minimum impact parameter may be affected, especially if the variations in the solar irradiance are small during the measurements [9], in which case better results are expected.

The effect of other bodies in the solar system can be evaluated from eq. 2. The most significant contributions are due to the Jupiter system (~ 163 cm), the Earth system (~ 30 cm) and the Saturn system (~ 25.2 cm). Table 2. reports the average Shapiro time delay, $\delta t_{1PN}$, produced by the planets in the solar system. The error due to the computation of the $\delta t_{1PN}$ in the perturbing body space-time frame instead of the solar system barycenter frame (i.e., the $\delta t_{SSB}$) is of the order of 0.02 mm (1-way) for Jupiter and at least an order of magnitude lower for other celestial bodies. $\delta t_{SSB}$ is negligible for planetary systems.

Table 2. Effect of relativistic correction, $\delta t_{1PN}$, due to the planets in the solar system applied to the first superior conjunction of BepiColombo.

| Celestial objects causing perturbation | $\delta t_{1PN}$ average value of relativistic correction |
|---|---|
| Mercury | 0.26 cm |
| Venus | 3 cm |
| Earth Barycenter | 15 cm |
| Mars Barycenter | 0.2 cm |
| Jupiter Barycenter | 162 cm |
| Saturn Barycenter | 25.2 cm |
| Uranus Barycenter | 2 cm |
| Neptune Barycenter | 1.5 cm |

We can conclude that for precise superior conjunction experiments with state-of-the-art radio tracking systems, it is recommended to adopt a full second order expansion of the light-time.

Eq. 4 and eq. 5 provide essentially the same values (as indicated above, the difference is at most 0.8 micron). For this reason, we implemented eq. 4 for the light-time computation in the MONTE navigation toolkit, as a compromise between accuracy and computational burden. We verified that the numerical results are equal to the analytical ones, shown above. In addition, we computed accordingly the adjustment to the partial derivative of $\gamma$, eq. 14. This correction is negligible and leads to no variation in the attainable formal uncertainty of $\gamma$.

$$\frac{\partial t_{2PN}}{\partial \gamma} = \frac{GM_\odot}{c^3} \ln\left(\frac{|\boldsymbol{r_2}| + |\boldsymbol{r_1}| + |\boldsymbol{r_{12}}|}{|\boldsymbol{r_2}| + |\boldsymbol{r_1}| - |\boldsymbol{r_{12}}|}\right) + \frac{G^2 M_\odot^2}{c^5} 2 \frac{|\boldsymbol{r_{12}}|}{|\boldsymbol{r_2}||\boldsymbol{r_1}|} \left[\frac{\arccos(\hat{\boldsymbol{r}}_1 \cdot \hat{\boldsymbol{r}}_2)}{|\hat{\boldsymbol{r}}_1 \times \hat{\boldsymbol{r}}_2|} - \frac{(\gamma + 1)}{1 + \hat{\boldsymbol{r}}_1 \cdot \hat{\boldsymbol{r}}_2}\right] \quad (14)$$

## 5. Conclusions

The remarkable advancements in the accuracy of radio-tracking instrumentation led us to question whether the Moyer implementation of the light-time was sufficiently accurate for current and future superior conjunction experiments. We performed numerical simulations to break down the different contributions to the light-time for the first superior solar conjunction of BepiColombo (see Table 1), which occurred between 10 and 24 March 2021. The Moyer approximation introduces an error of 17 mm (2-way) with respect to a complete second order expansion of the light-time. This is not negligible by MORE due to the novel PN ranging system @ 24 Mcps, which has an accuracy at centimeter level. We concluded that state-of-art radio tracking instrumentations are sensitive to a full second order expansion of the light-time. For present [11] and future superior conjunction experiments [43]–[45] it is recommended to use the accurate formulation of eq. 4, or eq. 5, while previous missions could safely rely on the Moyer formulation (eq. 3). The effects due to the Sun oblateness and angular momentum are not significant contributions. Furthermore, the contribution to the light-time of all planets of the solar system amounts to a total of $\sim$ 196 cm (2-way). In this case, the classical Shapiro time delay (eq. 2) is sufficient. For accurate applications, the perturbation to the light-time induced by planetary systems must be taken into account as well as the perturbation induced by the Sun.

Acknowledgments

The authors would like to thank N. Ashby, O. Luongo, D. Serra, G. Tommei and the colleagues of the Radio Science Laboratory of Sapienza University of Rome for useful discussions. This work was partially funded by the Italian Space Agency under the contract 2017-40-H.0.